\documentclass[aps,pra,twocolumn,showpacs,reprint,nofootinbib]{revtex4-1}
\usepackage{color,amsmath,amssymb,amsthm,times,graphics,graphicx,bm,dcolumn}

\newcommand{\be}{\begin{equation}}
\newcommand{\ee}{\end{equation}}
\newcommand{\ben}{\begin{eqnarray}}
\newcommand{\een}{\end{eqnarray}}
\newcommand{\bes}{\begin{subequations}}
\newcommand{\ees}{\end{subequations}}

\newcommand{\arxiv}[2][arxiv:]{\href{http://arxiv.org/abs/#1#2}{#1#2}}

\begin{document}

\title{Entanglement and entangling power of the dynamics in light-harvesting complexes}

\author{Filippo Caruso$^{1,2,3}$, Alex W. Chin$^{1,4}$, Animesh Datta$^{2,3}$, Susana F. Huelga$^{1,4}$, Martin B. Plenio$^{1,2,3}$}
\affiliation{$^1$ Institut f\"{u}r Theoretische Physik, Universit\"{a}t Ulm, Albert-Einstein-Allee 11, D-89069 Ulm, Germany}
\affiliation{$^2$ QOLS, The Blackett Laboratory, Prince Consort Road, Imperial College, London, SW7 2BW, UK}
\affiliation{$^3$ Institute for Mathematical Sciences, 53 Prince's Gate, Imperial College, London, SW7 2PG, UK}
\affiliation{$^4$ School of Physics, Astronomy \& Mathematics, University of Hertfordshire, Hatfield, AL10 9AB, UK}

\begin{abstract}
We study the evolution of quantum entanglement during exciton
energy transfer (EET) in a network model of the Fenna-Matthews-Olson
(FMO) complex, a biological pigment-protein complex involved in the
early steps of photosynthesis in sulphur bacteria. The influence of
Markovian, as well as spatially and temporally correlated (non-Markovian)
noise on the generation of entanglement across distinct
chromophores ({\em site} entanglement) and different excitonic eigenstates ({\em mode}
entanglement) is studied for different injection mechanisms, including thermal
and coherent laser excitation. Additionally, we study the entangling power
of the FMO complex under natural operating conditions. While quantum information
processing tends to favor maximal entanglement, near unit EET is achieved as the
result of an intricate interplay between coherent and noisy processes where
the initial part of the evolution displays intermediate values
of both forms of entanglement.
\end{abstract}

\maketitle

\date{\today}

\section{Introduction}
Photosynthesis, at its simplest, is the absorbtion of sunlight by
photosensitive antennae and its subsequent conversion into
chemical energy at a reaction center. The locations for these
processes are physically and physiologically separated, which
forces nature to devise a way for transferring the solar energy
from the antennae to the reaction center (RC). Exciton energy transfer
(EET) is facilitated by certain protein molecules called
light-harvesting complexes, and occurs at efficiencies of about
$99\%.$  EET has been a subject of continual interest for decades, not only for its phenomenal efficiency but also
for its fundamental role in Nature~\cite{forster48}.
Recently, ultrafast optics and nonlinear spectroscopy experiments
have provided new insights into the process of EET in
light-harvesting complexes like the one found in purple bacteria (LH-I) and the
Fenna-Matthew-Olson (FMO) complex~\cite{fleming07b,adolphs06}. In
particular, evidence of quantum coherence has been presented, with
the idea that nontrivial quantum effects may be at the root of its
remarkable efficiency~\cite{fleming07a}. Following this, several
studies have attempted to unravel the precise role of quantum
coherence in the EET of light-harvesting
complexes~\cite{aspuruguzik08,plenio08,ccdhp09,cdchp09,olaya,akifle,sarov},
and have, perhaps surprisingly, found that environmental decoherence
and noise plays a crucial role ~\cite{aspuruguzik08,plenio08,ccdhp09,cdchp09,chp}.

Light-harvesting complexes consist of several chromophores
mutually coupled by dipolar interactions residing within a protein
scaffold. Due to their mutual coupling, light-induced excitations on
individual chromophores (sites) can undergo coherent transfer from site
to site, and the typical eigenstates are therefore delocalized over
multiple chromophores. It is in these eigenstates, henceforth referred
to as exciton states, that one finds evidence of quantum coherence.
Here, we will study the role of quantum coherence in the process
of EET in the FMO complex, as quantified by quantum entanglement,
and investigate the sensitivity of the entanglement dynamics to
variables that have not been directly measured, such as the microscopic
interaction strength between the complex and the
surrounding environment, and the possible existence of spatial and
temporal correlations in the {\em bath}. In this context, the first analysis of the
entanglement behaviour in light-harvesting complexes was presented in
Ref. \cite{ccdhp09}, in which we analyzed
the evolution of an entanglement measurement, i.e.  logarithmic negativity
\cite{Logneg,PlenioV07}, in
a Markovian model of the FMO complex. The scope of the present work
is two-fold, on the one hand it aims to make the study of coherence
and entanglement in such systems quantitative by considering spatial and temporal (non-Markovian) noise correlations,
and secondly, it uses this quantitative approach to show that maximal entanglement
is not correlated with optimal transport, a result that may shed
light on the possible functional role of entanglement in EET.

Entanglement is defined between subsystems of a global system. When
considering entanglement in a composite
system whose components are closely spaced and strongly interacting,
as in the FMO complex, this choice of subsystems is, to some extent,
dictated by the way we interrogate the system. If the sites can be
addressed individually, then it is operationally well
justified to speak
about site-entanglement, i.e. quantum correlations across
distinguishable locations. However, if we are limited to accessing
the global excitations of the systems, i.e. the excitonic
eigenstates of the Hamiltonian governing the dynamics of the FMO
complex, then we will speak of mode entanglement as we then
explore entanglement between the eigenmodes of the system
(for more details on entanglement theory and its essential foundations see Ref. \cite{PlenioV07}).
Here, we will explore both types of correlations but place
a perhaps greater emphasis on the aspects of site-entanglement
which is more closely linked to the non-local structure of quantum
correlations. In general, the presence of quantum coherence, signified
by the presence of off-diagonal elements in the density matrix, is
necessary but not sufficient for the presence of quantum entanglement
\cite{PlenioV07}. However, the two conditions are equivalent
if one makes the idealized assumption that there is a single excitation in
the system. In this work, as explained below,
our entanglement analysis goes also beyond the single-excitation
approximation and takes into account different experimental and natural operating conditions
for the FMO complex dynamics.

The paper is organized as follows. In Sec. \ref{Site Entanglement} we introduce the theoretical
model for the FMO dynamics and discuss the entanglement measure used here. Then, we analyze the
logarithmic-negativity and the so-called entangling power in some non-Markovian FMO models in Sec. \ref{Non-Markovian models}
and generalize these results for different injection schemes, i.e. thermal injection and laser excitation of the FMO complex, in Sec. \ref{Beyond single excitons}.
Finally, the entanglement between excitons (mode entanglement) is investigated in Sec. \ref{Mode entanglement} and
the conclusions and final remarks are presented in Sec. \ref{Conclusions and Outlook}.
\section{Site Entanglement}
\label{Site Entanglement}
Each chromophore (site) in the FMO complex can be represented by a
two-state system (qubit). As in Ref. \cite{ccdhp09}, the effective
dynamics is modelled by an $N=7$ qubit
Hamiltonian which describes the coherent exchange of excitations
between sites, i.e.,
\be
\label{ham}
H = \sum_{j=1}^7 \hbar\omega_j \sigma_j^{+}\sigma_j^{-} + \sum_{j\neq l} \hbar v_{j,l} (\sigma_j^{-}\sigma_{l}^{+} + \sigma_j^{+}\sigma_{l}^{-}) \; ,        \ee
and local Lindblad terms that take into account the dissipation and dephasing caused by the surrounding environment, i.e.,
 \be
{\cal L}_{diss}(\rho) = \sum_{j=1}^{7} \Gamma_j[-\{\sigma_j^{+}\sigma_j^{-},\rho\} + 2 \sigma_j^{-}\rho \sigma_j^{+} ]
\ee
and
\be
{\cal L}_{deph}(\rho) = \sum_{j=1}^{7} \gamma_j[-\{\sigma_j^{+}\sigma_j^{-},\rho\} + 2 \sigma_j^{+}\sigma_j^{-}\rho \sigma_j^{+}\sigma_j^{-}] \; .
        \ee
This Markovian form of the evolution preserves complete positivity, an essential feature when evaluating entanglement, as discussed below. Here $\sigma_j^{+}$ ($\sigma_j^{-}$) are the raising and lowering operators for site $j$, $\hbar\omega_j$ is the local site excitation energy, $v_{k,l}$ denotes the hopping rate of an excitation between the sites $k$ and $l$, and $\Gamma_j$ and  $\gamma_j$ are the dissipative and dephasing rate at the site $j$, respectively. Finally, the transfer efficiency is measured in terms of an irreversible transfer of excitations (with rate $\Gamma_{sink}=6.283~\mathrm{ps}^{-1}$) from site $3$ to an extra site $8$ modeling the RC, as described by the Lindblad operator
\be
 {\cal L}_{sink}(\rho) = \Gamma_{sink}[2\sigma_{8}^{+}\sigma_3^{-}\rho \sigma_3^{+}\sigma_{8}^{-} - \{\sigma_3^{+}\sigma_{8}^{-}\sigma_{8}^{+} \sigma_3^{-},\rho\} ] \; .
\ee
Particularly, the transport efficiency is described by the population transferred to the sink $p_{sink} (t)$, which is given by
 \be
p_{sink}(t) = 2\Gamma_{sink}\int_{0}^t p_{3}(t')\mathrm{d}t' \; ,
 \ee
where $p_3(t')$ is the population of the site $3$ at time $t'$. Moreover, we choose $\Gamma_j = 5 \times 10^{-4}~\mathrm{ps}^{-1}$ for any site $j$, as in Ref. \cite{ccdhp09}. Notice that decoherence appears in the model above via the action of a pure-dephasing Lindblad super-operator in the master equation for the exciton dynamics, which is equivalent to having stochastic fluctuations of the exciton site energies induced by the environment. This model is known as the Haken-Strobl model and has been used extensively in the chemical physics literature to describe exciton dynamics over several decades \cite{HS}. In the next section, we will investigate the entanglement behaviour in some non-Markovian models of the FMO complex dynamics with different types of environmental interactions.

Let us stress that, in order to obtain sensible and reliable results for the entanglement analysis, it is crucial that the exciton dynamics is represented by a completely positive map. Indeed, we numerically found that small deviations from the complete positivity conditions are enough to cause significant changes to the entanglement. The Lindblad formalism absolutely guarantees the positivity of the evolving state, whereas the majority of non-perturbative and non-Markovian treatments do not. Hence, in the following, we deem it prudent to choose and to analyze only non-Markovian models, that give both reliable entanglement predictions and which are consistent with the essential experimental transport data. Actually, very little is actually known about the microscopic details of the environment in FMO complex, and it is very hard to distinguish between the various noise models proposed in literature -- see for instance Refs. \cite{akifle,adolphs06,sharp,scw} -- using the available experimental data.
 One may conclude that the additional complexity of all of these models results from the introduction of a range of new variables which can be all independently tuned to match the key experimental results, even though the detailed dynamics of these models may be quite different. It should be noted in this respect that the simple Lindbladian model we use here can in fact account for the key features of the experimentally
observed dynamics, i.e. long-lived coherence times and transport times -- see Ref. \cite{ccdhp09}.

We will quantify the entanglement across a bipartition $A|B$ of a composite system by using the logarithmic negativity \cite{PlenioV07}, i.e.
 \be
E(A|B)=\log_2||\rho^{\Gamma_A}||_1 \; ,
\ee
where $\Gamma_A$ is the partial transpose operation of the density operator $\rho$ with respect to the subsystem $A$ and $|| \cdot ||_1$ denotes the trace norm. It quantifies how negative the spectrum of the partial transpose of the density matrix is, consequently it is only meaningful if the evolution is completely positive. Furthermore,
we would like to stress that in addition to its computational simplicity, the
logarithmic
negativity also possesses an operational interpretation in terms of the entanglement cost for the exact preparation of the state \cite{AudenaertPE}.
When confined to the one excitation subspace, if $A=1 \dots k$ is a set of $k$
chromophores within a global system of $N$ sites, then the logarithmic negativity
across the bipartition $(1 \dots k)|(k+1 \dots N)$ is given by the compact expression
\be
E(1
\dots k|k+1 \dots N)=\log_2\left( 1-a_{00}+\sqrt{a_{00}^2 + 4 X} \right)
\ee
where $X=\sum_{i=1}^k \sum_{j=k+1}^N |a_{ij}|^2,$ and $a_{ij}$ denotes the
off-diagonal element between states with  excitations in qubits
$i$ and $j$. Here $a_{00}$ is the matrix element corresponding to the zero excitation subspace. If all coherences are vanishing, i.e. $a_{ij}=0$, there is no entanglement across any partition in the one-excitation sector. Note that the restriction to at most a single excitation is
unproblematic in the case of the logarithmic negativity as it does not affect
its definition and the fact that it is an entanglement monotone, that is
non-increasing, for general local operations and classical communication (LOCC).
This is not the case if the
constraint to at most a single excitation is applied in a way that amends
the definition of the functional. In Ref. \cite{fleming}, for example, the global
entanglement $E_G$, which is defined as the relative entropy of entanglement
with respect to the set ${\cal S}$ of totally separable states \cite{relent},
was amended to yield $E_{G1}$ by replacing the set ${\cal S}$ of totally
separable states by the set ${\cal S}_1$ of totally separable state with at
most one excitation. This new function $E_{G1}$ is non-increasing only when
we restrict attention to the set of LOCC operations ${\cal O}_1$ that map
${\cal S}_1$ into itself. The set ${\cal O}_1$ excludes a wide range of
important physical processes such as non-diagonal local unitaries and hence
fundamental processes such as laser excitation because the raising operator $a^{\dagger}\notin {\cal O}_1$ does not map
${\cal S}_1$ onto itself. Hence, while being computable $E_{G1}$ is not an
entanglement monotone under natural, physically realizable and important
operations that are routinely applied. In fact, the operations permitted
in ${\cal O}_1$ do not create coherence between the zero and the single
excitation sector and hence a more natural interpretation of $E_{G1}$ is one of quantifying
coherence rather than entanglement.

Therefore we will restrict attention to the logarithmic negativity, which
is an accepted and at the same computable entanglement monotone ~\cite{PlenioV07} for arbitrary LOCC
and arbitrary excitation levels.

\section{Non-Markovian models}
\label{Non-Markovian models}
To estimate the impact of non-Markovian effects on the entanglement
in the FMO complex, we will consider two non-Markovian models of the
FMO complex dynamics, corresponding to a different type of interaction
with the surrounding environmental. In particular, we assume the FMO
complex to be linearly coupled to a bath of damped harmonic oscillators.
In all cases, the harmonic oscillators will be damped into a zero
temperature bath at rate $\kappa$. By using these non-Markovian models,
which manifestly preserve the completely positivity of the corresponding
quantum evolution, we can study the behaviour of the entanglement,
measured by the logarithmic negativity and by the entangling power of the
quantum evolution itself.
\subsection{Local bath model}
While the phenomenon of EET is clearly noise-assisted, the exact dynamics
within the observed exciton transmission time are strongly model-dependent.
To further emphasize this aspect and therefore invoking the need of further
experimental results, we will present in this section a tunable
noise model which reproduces the Markovian results presented in \cite{ccdhp09}
in certain parameter regime but, interestingly enough, can also provide longer
coherence times while preserving and even enhancing EET in a parameter regime
where the model exhibits a degree of non-Markovianity. In particular, we consider
an environment model motivated by the approach presented by Adolphs and Renger
in Ref.~\cite{adolphs06}, where sites interact locally with a quasi-resonant
localized mode. Their spectral density contains a contribution from a low-energy
continuous density of states and a discrete high-energy mode, and its effects on
the dynamics of a dimer molecule were recently simulated using a new application of
the time-adaptive renormalisation group method~\cite{prior}. Here, for simplicity,
we will consider a model in which each FMO chromophore is linearly coupled to a
resonant harmonic mode with strength $g$ while each mode is damped into a zero
temperature Bosonic reservoir with strength $\kappa$. In order to describe these
couplings, we add to the previous Hamiltonian in Eq. (\ref{ham}) the following
two terms
\ben
H_B &=& \sum_{j=1}^7 \hbar\omega_h^j a_j^{+} a_j \; , \\
H_{SB} &=& \sum_{j=1}^7 g_j (a_j + a_j^{+}) \sigma_j^{+}\sigma_{j}^{-} \; ,
\een
where $H_B$ is the free Hamiltonian for the two-level bath with creation and annihilation operators $a^{+}$ and $a$, respectively, and mode frequency $\omega_h^j=\omega_j$, and $H_{SB}$ is the system-bath interaction Hamiltonian with interaction strength $g$. The damping is introduced by considering a Lindblad term ${\cal L}^{l-bath}_{rad}(\rho)$ of the form
\be
{\cal L}^{l-bath}_{rad}(\rho) = \sum_{j=1}^{7} \kappa_j
[ -\{ a_j^{+} a_j ,\rho\} + 2  a_j \rho a_j^{+} ]
\ee
with $k_j$ being the rate at which the local harmonic mode, coupled to the site $j$, is damped
into a zero temperature bath. Within the considered parameter regimes, the local modes can be reasonably considered within a two-level approximation. In fact, we numerically monitor the populations in each local mode and we found that no local modes were strongly excited or saturated over the whole time interval investigated here.
\begin{figure}[t]
\centerline{\includegraphics[width=.49\textwidth]{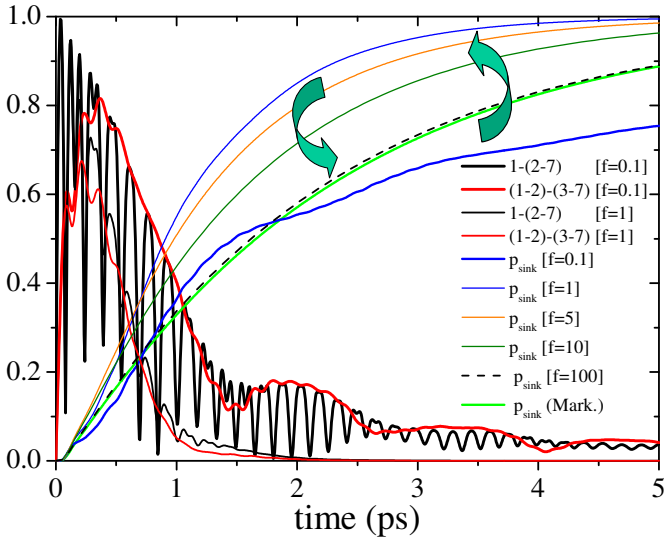}}
\caption{Entanglement of different bi-partitions of the FMO
complex dynamics subject to a local form of noise where each FMO site couples to a damped resonant mode. The effective
dephasing rates are taken to be equal to the optimal values derived for a fully Markovian model as in Ref. \cite{ccdhp09} but
the degree of Markovianity of the model can be tuned by varying the parameter $f$, as described in the text.
The entanglement curves are for the 6 splits of the form $(1,\cdots,i)|(i+1,\cdots,7)$, $i \in\{1,\cdots,6\}$, where
$i$ corresponds to the FMO site $i$. The behaviour of $p_{sink}$ for different values of $f$ is also shown. Large values of $f$ render the dynamics fully Markovian and we recover the results in Ref. \cite{ccdhp09}, while one can observe longer coherence times while at the same time enhancing $p_{sink}$ in the low $f$ domain, as the one exemplified by the value $f=1$. Note the non-monotonic behaviour of the transfer rate as a function of $f$, as emphasized by the arrows.}\label{fig1b}
\end{figure}
The damping rate $\kappa_i$ determines the width of the spectrum and
hence the
correlation time of the local environment associated to the site $i$. For weak coupling and strong mode-losses,
this approach leads to a Markovian environment with Lorentzian line shape (similar
to the models studied by different methods in~\cite{Ishizaki,prior}), while, for low
losses and strong coupling, the high degree of excitation of the environment
leads to deviations from the Lorentzian lineshape. To isolate the impact
of the non-Markovianity, we keep the ratio $g^2/\kappa$, i.e. the effective
coupling strength between the site and its mode, fixed while varying the
ratio $g/\kappa$. To this end, we employ a parameter $f$ to parametrize the
system-mode coupling rates $g=\sqrt{f} g_0$ and energy loss rates
of the modes with $\kappa = f \kappa_0$. Then for $f\gg 1$ ($g\ll \kappa$),
i.e. the Markovian limit, we reproduce the optimized dephasing rates found in
\cite{ccdhp09}, while for $f\ll 1$ ($g\gg \kappa$) we find non-Markovian behaviour.
We initiate the system with one excitation in site 1. For a single site this
dynamics has been tested to reproduce both the correct Markovian limit and
to be capable of exhibiting strongly non-Markovian behaviour as illustrated in Fig. \ref{fig1b}.
Here, we choose the system-mode coupling
rates to be equal to $g_0=\kappa_0=\{
1,50,41,50,41,5,50\}/5.3~\mathrm{ps}^{-1}$ to match closely the effective dephasing
rates $\{0.157,9.432,7.797,9.432,7.797,0.922,9.433
\}~\mathrm{ps}^{-1}$ in the Markovian limit, as in Ref. \cite{ccdhp09}.
In Fig. \ref{fig1b}, we show the entanglement behaviour for two different bi-partitions of the FMO
complex and for two values of $f$ far from strict Markovianity as well as the time-dependence for $p_{sink}$ when $f$ varies from $f=0.1$ to the value of $100$. The behaviour obtained in the fully Markovian case of Ref. \cite{ccdhp09} is also represented and is indeed already reached for $f=100$. Interestingly, we obtain that taking $f$ as our measure of deviation from Markovian behaviour, $p_{sink}$ varies non-monotonically with $f$ so that, differently from the non-local bath non-Markovian model below or the model described in \cite{Ishizaki}, here the presence of a degree of temporal correlations may even assist the transport of electronic excitations from the antenna to the RC, as seen for values of $f$ in the range $1-10$. On the other hand, very low values of $f$ (strong non-Markovianity) lead to decreased transport while preserving large values of the coherence across site bipartitions. Note that different noise models, as the one presented in \cite{rcg}, yield different conclusions.
This also shows the great uncertainty about the nature of transport dynamics that arises from our ignorance of the microscopic details of the environment. In this respect, experimental entanglement measurement could be important for removing some of this uncertainty, as it can be very sensitive to the structure of the noise the system is being subject to \cite{taka}.
\subsection{Non-local bath model}
\begin{figure}[t]
\centerline{\includegraphics[width=.49\textwidth]{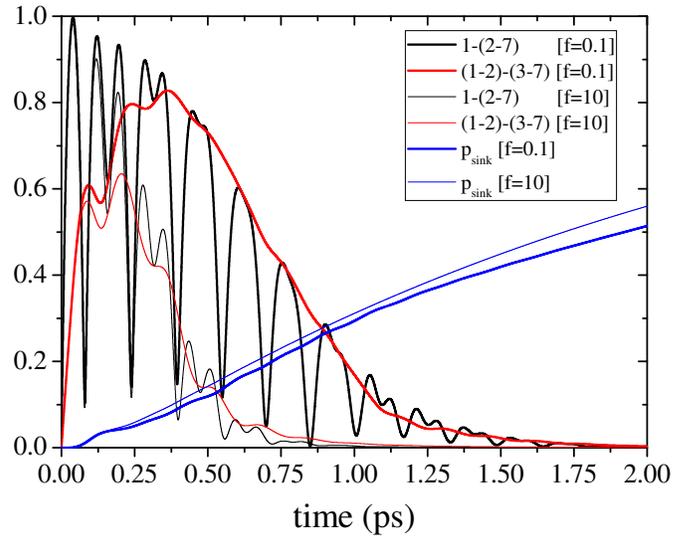}}
\caption{Entanglement of different bi-partitions of the FMO
complex vs. time ($\mathrm{ps}$), in the presence of a tunable source of noise correlations described
by a non-local Bosonic bath,
for different values of the ratio $f$. For
small $f$, the environmental correlations yield a non-Markovian form
of noise and the time scale for entanglement persistence increases while, within this model, the efficiency for
transport $p_{sink}$ decreases.}\label{fig1a}
\end{figure}
In this subsection we would like to go further and explore more sophisticated
entanglement properties of the entanglement dynamics including its entangling
power. To this end we need to further simplify the decoherence model to permit
its numerical analysis in the context of the entangling power. This model will
also include non-local correlations in the environment and we will start by
considering again the entanglement of states in the FMO dynamics under this
model and then apply it to the study of the concept of entangling power.
\begin{figure}[t]
\includegraphics[width=.49\textwidth]{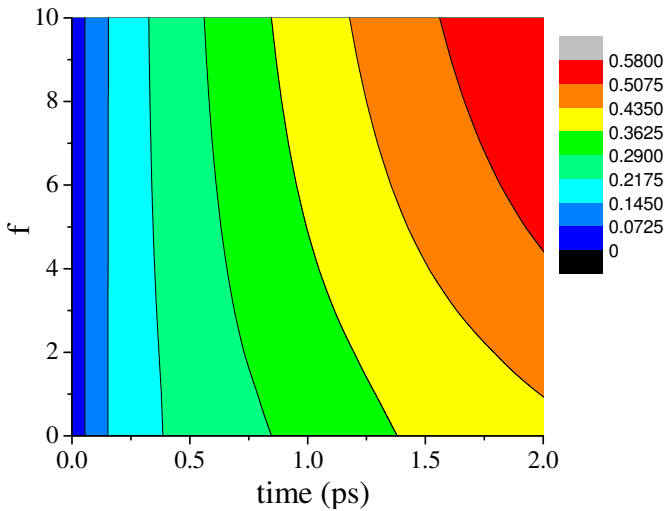}
\caption{Contour plot for the transfer efficiency in FMO complex, as a function of time (in $\mathrm{ps}$) and of the
parameter $f$, in the case of a non-Markovian model with a non-local
Bosonic bath linearly interacting with the FMO.}\label{fig2a}
\end{figure}

Formally, we consider a model in which we couple each site linearly, with
strength $g_j$, to a single, common harmonic mode damped
into a zero temperature bath at rate $\kappa_j$ that depends on the location of
the excitation \cite{note}. In particular, the following bath-system interaction Hamiltonian term $H_{SB}$ is added to previous
Hamiltonian in Eq. (\ref{ham}), i.e.
\be
    H_{SB} = \sum_{j=1}^7 g_j (a + a^{+}) \sigma_j^{+}\sigma_{j}^{-} \; ,
\ee
and the damping is described by the following Lindbladian super-operator
\be
    {\cal L}^{g-bath}_{rad}(\rho) = \sum_{j=1}^7 \kappa_j [ -\{ \mathbb{P}_j a^{+} a \mathbb{P}_j,\rho\} + 2   \mathbb{P}_j a \rho  a^{+} \mathbb{P}_j] \;,
\ee
where $a$ and $a^{+}$ are, respectively, the annihilation and creation operators
of the harmonic mode and $\mathbb{P}_j$ is the projector onto site
$j$ in the FMO complex. Hence the line width of the harmonic oscillator will
depend on the location of the excitation.
This model mimics closely the model of the previous section in that sites see
harmonic oscillators with site--dependent damping rates. At the same time, the
Hilbert space of this model is much smaller as the dynamics is restricted
to the single excitation sector only. In effect we can describe the dynamics in
the basis $\{|i\rangle|0\rangle,\ldots,|i\rangle|d\rangle\}_{i=1,\ldots,7}$ where
the first index refers to the excitation in the FMO complex and the second refers
to the environment oscillators. The damping rate of the environment oscillator
can be made to depend on the site in which the excitation resides in as would be
the case in the local model. Hence this model represents a mix between the local
model in the previous subsection and that of an FMO complex coupled to a single
mode the former allowing for different coupling rates and line widths for
different sites and the latter allowing for a significant reduction in the
Hilbert space dimension of the simulation.
In Fig.~\ref{fig1a}, we consider the entanglement between site $1$ and
the remainder of the FMO complex as well as sites $1$ and $2$ versus the
remainder of the FMO complex.  Here, we choose the system-mode coupling
rates as in the local non-Markovian model above, and we find that
non-Markovian effects (decreasing $f$)
reduces the transport efficiency while it prolongs the lifetime of entanglement.
These observations are explained by the fact that the non-Markovian
dephasing leads to a reduction of the effective noise level
in the system (a phase flip by the environment may be followed
by another correlated phase flip at a later time, hence canceling
out), upsetting the optimal balance of quantum and incoherent dynamics
required for efficient EET \cite{ccdhp09,cdchp09}, whilst
also increasing and
preserving the entanglement that is present in the system. The fact that
the presence of a non-Markovian environment may enhance entanglement
content beyond the values predicted for an evolution subject to
memoryless environments was also predicted when analyzing strictly
bipartite systems \cite{palermo,durban}, including biological scenarios
\cite{reina}.
\begin{figure}[t]
\includegraphics[width=.49\textwidth]{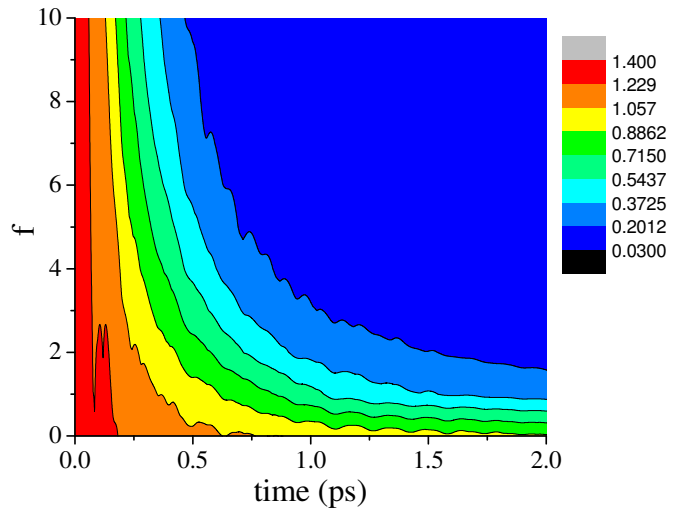}
\caption{Contour plot for the entangling power in FMO complex, as a function of time (in $\mathrm{ps}$) and of the
parameter $f$, in the case of the non-local bath non-Markovian model. For
low noise levels the entangling power of the evolution can be
non-monotonic in time, indicating coherent oscillations in the
system. The entangling power decreases monotonically with increasing
noise level. As comparison, the entangling power of a CNOT gate is 1, and a SWAP gate is 2.
This means that the FMO complex has nontrivial entangling power for a large parameter range.}\label{fig2b}
\end{figure}
Moreover, rather than exploring the entanglement content of states that
are generated during the evolution, one may also study quantitatively
the entanglement content of the quantum evolution itself. Just as
for quantum states such a quantification is, however, not unique.
Here, we consider the capacity of an evolution to create entanglement
between two subsystems, each of which may be composed of several
components. This automatically provides a lower bound
for the amount of entanglement that is required to reproduce the
dynamics of the system purely from local operations and classical
communication \cite{entpow1,entpow2}. In
Figs.~\ref{fig2a} and \ref{fig2b}, we show a contour
plot for the transfer efficiency and the entangling power
of the noisy evolution of the FMO complex in
the first picoseconds of the EET process, respectively.
In particular, we consider the entangling
power for the evolution between different bi-partitions of the
system, quantified by the logarithmic negativity, in the FMO when initially prepared in
a maximally entangled state with $7$ ancilla qubits. We consider
the split \{qubits $1_{FMO}$--$1_{ancilla}\}$--\{the rest\},
as a function of time (in $\mathrm{ps}$) and of the parameter above $f$.
Large quantum correlations are not associated with optimal transport.
In the absence of any dephasing, entanglement lasts for
times limited only by the excitation loss rate. On the other hand, non-Markovian
dephasing (decreasing $f$) increases the entanglement,
which persists for about the initial $20\%$ of the total
transmission time, but decreases transport efficiency.
%
%
%
%%%%%%%%%%%%%%%%%%%%% beyond single excitations %%%%%%%%%%%%%%%%%%%%%%%%%
\section{Beyond single excitons}
\label{Beyond single excitons}
So far we have assumed that the system is initialised with a
single excitation in site 1. This may not be realised precisely
under experimental or natural operating conditions and, furthermore,
neglecting higher excitations (though only existing for a really short time) may influence the entanglement
content of the system considerably.
Hence, a study of quantum entanglement in the FMO complex
under realistic
conditions should consider a model which allows the freedom to
control the number of excitations in the complex at any time. To
this end, we generalize the theoretical noise model proposed in \cite{plenio08,ccdhp09},
by first modeling (i) the baseplate feeding excitations into
the FMO complex as a thermal reservoir of excitations at an
effective temperature $T$ and then (ii) a system under laser pulse
irradiation.
\subsection{FMO thermal injection}
\begin{figure}[t]
\centerline{\includegraphics[width=.5\textwidth]{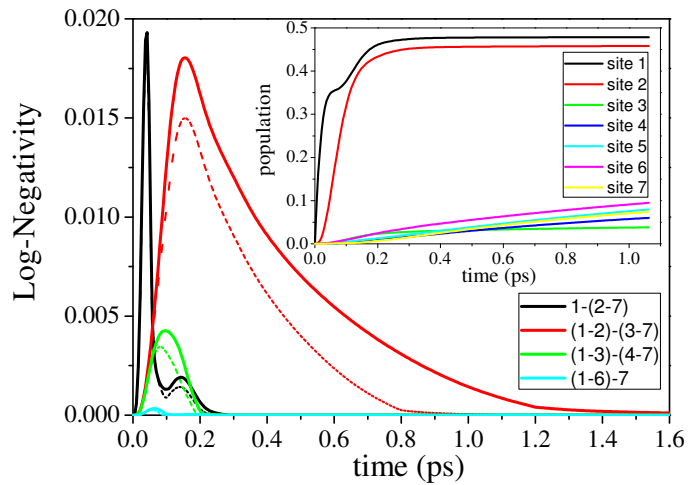}}
\caption{Entanglement, quantified by the
logarithmic negativity, in the FMO complex for local (dashed
lines) and for spatially correlated (continuous lines) dephasing noise, in presence
of the thermal injection bath ($n_{th}=100$, $\Gamma_1=1$), within the Markov approximation.
The inset shows the site population behaviour as a function
of time.}\label{mixing}
\end{figure}

The complex starts in the ground state, without any excitations,
which are introduced into the network via the site $i$ with a rate $\Gamma_i$. This
process is modelled by a thermal bath of harmonic oscillators at a
temperature given by the thermal average boson number $n_{th}.$
Within the Markov approximation, the Lindblad superoperator for
the injection of excitations assumes the form
 \ben
 {\cal L}_{inj}(\rho) &=& n_{th}\frac{\Gamma_{i}}{2}[-\{\sigma_{i}^{-}\sigma_{i}^{+},\rho\} + 2 \sigma_{i}^{+}\rho \sigma_{i}^{-}]
\nonumber \\ &+&(n_{th}+1)\frac{\Gamma_{i}}{2}[-\{\sigma_{i}^{+}\sigma_{i}^{-},\rho\} + 2 \sigma_{i}^{-}\rho \sigma_{i}^{+}] \; ,
 \een
which can now be used to study the evolution of quantum
entanglement in the FMO complex (or any light-harvesting complex)
under various possible natural settings. In particular, motivated by the experimental
observations, we choose the site $1$ as
the one at which the excitations are introduced in the FMO complex.
In Fig.~\ref{mixing},
we show the entanglement time evolution for various
subsystems of the qubit network modelling the FMO complex. As can
be seen, the amount of entanglement is considerably smaller
compared to when the FMO complex starts with \emph{exactly} one
excitation, say on site 1, although it persists on the same
timescale~\cite{ccdhp09}. This is because the second term in the
injection Liouvillian, which allows for an irreversible loss of
excitations from the complex (comparable to spontaneous decay),
always accompanies the first one which introduces excitations into
the complex.
Besides, given the dimensions and structure of the FMO complex (closest
site-site distance is $\sim 11 \mathrm{\AA}$), dephasing may not
be local, but correlated in space and time
\cite{plenio08,ccdhp09,correla,correla2}. For this reason, we consider also the case
of spatially correlated dephasing by using the following Lindblad term:
 \be
{\cal L}^c_{deph}(\rho) =- \sum_{m,n} \gamma_{mn}[A_m,[A_n,\rho]] \; ,
 \ee
where $\gamma$ is a positive
semidefinite matrix (to preserve the complete positivity of the quantum evolution), but with the diagonal
elements equal to the optimal local dephasing rates as in Ref. \cite{ccdhp09}, and $A_m =
\sigma_m^+\sigma_m^-$. Here, the amount and duration of entanglement
are both slightly enhanced (solid lines in Fig.~\ref{mixing}).
\subsection{FMO laser excitation}
\begin{figure}[t]
\centerline{\includegraphics[width=.49\textwidth]{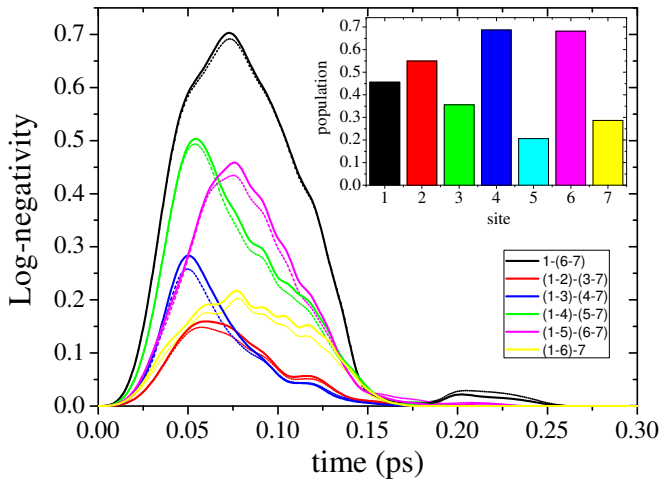}}
\caption{Entanglement in FMO complex (evolving by a Markovian dynamics) when irradiated with a
laser pulse (see the text for more details).
Dephasing noise is either local (dashed lines) or spatially
correlated (continuous lines). Strong entanglement
between sites $1$ and $2$ is clearly illustrated by the behaviour observed
in the top curve (bipartition $1-(2-7)$) as opposed to the bottom one
(bipartition $(1-2)-(3-7)$). Similarly, strong entanglement between
sites $5$ and $6$ is best exemplified by the central plot where we compute the entanglement
across bipartition $(1-5)-(6-7)$. Inset: site population distribution
in the FMO complex for $t=75~\mathrm{fs}$.}\label{fig4}
\end{figure}

In the laboratory~\cite{fleming07a,fleming07b}, the complex
is typically irradiated with a short laser pulse centered on the typical transition frequencies
of the sites. The coupling between the FMO complex and
the external radiation field can be described by the semiclassical
time-dependent Hamiltonian $H_{FMO-laser}(t)$, which reads in rotating wave approximation as follows
 \be
 H_{FMO-laser}(t) = - \sum_{i=1}^{7} \vec{\mu}_i \times \vec{e} \ E(t) \ e^{-i \omega_1 t} \ \sigma^{+}_i + h.c.
 \ee
with $\vec{\mu}_i$ being the molecular transition dipole moment of the individual site $i$ (taken from the published crystal structure in Ref. \cite{tronrud}),
$\vec{e}$ being the polarization of the field, and $E(t)$ the time-dependent electric field. In particular, following Ref. \cite{adolphs06}, we consider a Gaussian electric field pulse
of width $60~\mathrm{fs}$, centered at $120~\mathrm{fs}$, with an electric
field strength $E_0=4.97968$ $D^{-1}$ $cm^{-1}$, polarized
parallel to the dipole moment of site 1, and with a frequency on resonance with the optical transition of site $1$ ($\omega_1)$.
The electric field amplitude was chosen to excite one excitation on site $1$, i.e. to give a $\pi$-pulse on site $1$.
However, as the molecular transition dipole moments are of the order of the intersite energy difference, the laser pulse leads to excitation of all sites.
We have studied this scenario as well, both for
local and correlated spatial dephasing noise, and the results are
presented in Fig.~\ref{fig4}. This scenario generates a
considerable amount of entanglement (though for a shorter amount
of time compared to the previous scenario), and it might be
concluded that although an FMO complex operating in nature may not
possess substantial amounts of entanglement, it is possible in a
laboratory to generate large amounts of it.
This could open up new
vistas for exploration of quantum effects in biological systems,
albeit under laboratory conditions, and also allow for a
demonstration of entanglement enhancement under non-Markovianity.
\section{Mode entanglement}
\label{Mode entanglement}
\begin{figure}[t]
\centerline{\includegraphics[width=.49\textwidth]{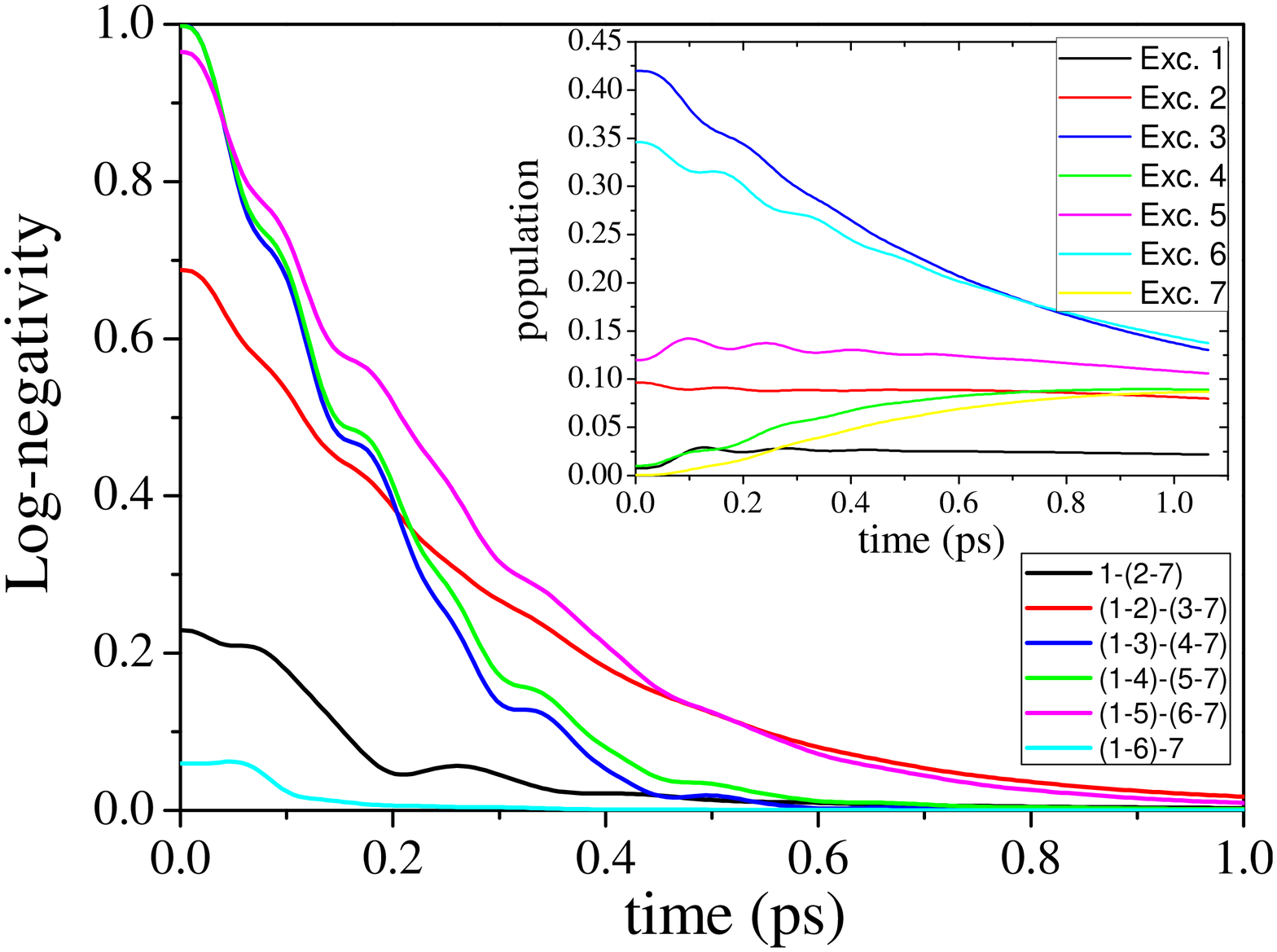}}
\caption{The logarithmic negativity in the FMO
complex (evolving by a Markovian dynamics) for local dephasing noise in the exciton basis.
Initially, one excitation is in the site $1$. The curves are
for the 6 splits of the form $(1,\cdots,i)|(i+1,\cdots,7)$, $i
\in\{1,\cdots,6\}$, where now $i$ corresponds to the exciton $i$
and they are ordered with increasing exciton energies. Inset:
exciton population vs. time ($\mathrm{ps}$).}\label{fig5}
\end{figure}
When local addressing is unfeasible, site entanglement is directly
immeasurable even when the evolution is confined to the single
excitation sector. However, the evaluation of mode entanglement
via the experimental determination of coherences in the exciton
basis can provide us with information about the existence of
quantum correlations in the system. The temporal behaviour of mode
entanglement within our Markovian model \cite{ccdhp09} and for different
bi-partitions is shown in Fig.~\ref{fig5}, together with
the exciton populations along the first $\mathrm{ps}$ of the EET. Initially
all modes are populated, with the largest fraction in excitons 3
and 6, which leads to high values of mode entanglement across
bipartitions each containing one of those high energy excitons. As
time elapses, entanglement degrades monotonically
as the transfer efficiency increases.
\section{Conclusions and Outlook}
\label{Conclusions and Outlook}
Efficient EET in light-harvesting complexes can be traced back to
an interplay between coherent and incoherent processes where the
quantum correlations characteristic of the coherent evolution are
partially suppressed by noise, yet not entirely destroyed. We have
placed the analysis of entanglement in such systems on a
quantitative footing and showed that for optimal transport the
entanglement, while present, is neither maximal nor long lived.
Actually, long lived entanglement exists in the absence of
dephasing which is known to be highly inefficient. However, despite the fact that
observed transfer times do require a noise-assisted transport dynamics,
it turns out that
the time-dependence of both coherences and population transfer over the full transfer time are strongly model-dependent.

In summary, an
interplay between creation of entanglement for short distances and
times (through coherent interaction) followed by the destruction
of entanglement for longer distance and times (through dephasing
noise) seems to be necessary for optimal transport. Moreover, our entanglement
results could actually be seen as providing a potential experimental test for
the form of the system-environment coupling. Unlike the coherence and transport times,
the entanglement is sensitive to the precise evolution of the system, and an experiment that could measure entanglement
(such experiments are indeed planned), could differentiate between models.
In order to do so, the entanglement predictions for various noise models must be available for comparison,
and in this respect the nature of the environment is central to the problem of energy transport in photosynthetic complexes. Further
studies are however required before any result can be accepted as conclusive
on the functional and possibly beneficial role of coherence and
entanglement in EET.

\section*{Acknowledgments}
This work was supported by the EPSRC, the EU projects QAP and CORNER,
and a Alexander von Humboldt Professorship.
A.W.C. is most grateful to G.R. Fleming and his group for their hospitality.
F.C. was supported also by
a Marie Curie Intra European Fellowship within the 7th
European Community Framework Programme.


\begin{thebibliography}{99}
\bibitem{note} A finite temperature setting may also be implemented
by initializing the mode in a thermal state at temperature $T$ and
coupling the mode to a Markovian environment at temperature $T$.
\end{thebibliography}
\end{document}